\documentclass[prb,twocolumn,showpacs]{revtex4}

\usepackage{bm}
\usepackage{graphicx}
\usepackage{times}

\newcommand*{\be}{\begin{equation}}
\newcommand*{\ee}{\end{equation}}
\newcommand{\ba}{\begin{eqnarray}}
\newcommand{\ea}{\end{eqnarray}}

\begin{document}

\title{Possible magnetic-field-induced voltage and thermopower in diluted
magnetic semiconductors}

\author{Carsten Timm}
\email{ctimm@ku.edu}
\affiliation{Department of Physics and Astronomy, University of Kansas,
Lawrence, KS 66045, U.S.A.}

\date{\today}

\begin{abstract}
In diluted magnetic semiconductors, the carrier concentration
and the magnetization of local moments are strongly coupled, since the magnetic
interaction is mediated by the carriers. It is predicted that this coupling
leads to an electric polarization due to an applied magnetic-field gradient and
to the appearance of a magnetic-field-dependent voltage. An expression for
this voltage is derived within Landau theory and its magnitude is estimated for
(Ga,Mn)As. Furthermore, a large contribution to the thermopower
based on the same mechanism is predicted.  The role of fluctuations is also
discussed. These predictions hold both if the magnetization is uniform and if
it shows stripe-like modulations, which are possible at lower temperatures.
\end{abstract}

\pacs{
75.50.Pp, 
75.80.+q, 
72.15.Jf, 
75.30.Fv} 

\maketitle

\section{Introduction}

Among magnetically ordered materials, diluted magnetic
semiconductors\cite{Ohn98,ZFS04,JSM06} (DMS) are unique in the possibility of
tuning the interaction between local moments \textit{in situ}. Several
groups\cite{OCM00,BKB02,NKS03} have demonstrated that the magnetization and
Curie temperature $T_c$ in DMS can be changed by applying a gate voltage in a
field-effect-transistor geometry. The gate potential affects the
concentration of carriers, usually holes, in the DMS sample, which leads to a
change of the magnetic interactions, since these are mainly mediated by the
carriers. A strong dependence of the magnetization on carrier concentration is
also seen in experiments on series of samples with different dopand
concentrations.\cite{EWC02,LLD05,KHT05}

On a microscopic level, each local moment magnetically polarizes the carriers
in its vicinity due to the exchange interaction between carriers and localized
(usually d-shell) electrons of impurity ions. The other local moments see these
carrier spin polarizations, leading to an effective magnetic interaction
between local moments. This is the essence of the Ruderman-Kittel-Kasuya-Yosida
(RKKY) interaction.\cite{RKKY} It is here complicated by band-structure effects
and spin-orbit coupling in p-type DMS,\cite{ZaJ02,BrG03,TiM04,KTD04} screening
due to disorder,\cite{PHS04} the small Fermi energy compared to the effective
Zeeman splitting,\cite{MZS04} and competing short-range interactions. The
RKKY-type interaction depends on carrier concentration essentially through the
density of states at the Fermi energy. This dependence can be expanded to
linear order in the concentration, in agreement with experimental
observations.


The present author has recently suggested that this coupling may stabilize an
equilibrium state with  periodic, but strongly anharmonic modulations of
carrier concentration and magnetization below a certain temperature $T^\ast
<T_c$.\cite{Tim06} The temperature $T^\ast$ strongly depends on $\eta \equiv
\partial T_c/\partial n$, the change of $T_c$ with carrier concentration. In
Landau theory we find $T_c-T^\ast\propto 1/\eta^2$. While some experimental
results\cite{LLD05} are consistent with a small value of $T_c-T^\ast\sim
10\,\mathrm{K}$, which would make the effect observable, it is not clear
whether this value is typical for DMS. In the present paper we put the main
emphasis on the uniform phase.

In particular, we explore the reverse effect, compared to the gate control of
magnetism: Can a change in magnetization lead to a change in carrier
concentration? Or, if the sample is electrically isolated, can it lead to the
appearance of a voltage? Microscopically the dependence of the effective spin
interaction on the carrier concentration means that there is an additional
contribution to the carrier energy depending on spin orientations. In
particular, if the impurity spins are partially aligned, there is an
\emph{attractive} potential for carriers. This effect is discussed and its
magnitude is estimated in Sec.~\ref{sec.magvolt}. In addition, the
magnetization of course also changes with temperature, which should generate a
contribution to the thermopower. This possibility is discussed in
Sec.~\ref{sec.thermop}. The paper is summarized in Sec.~\ref{sec.sum}.

\section{Magnetic-field-induced voltage}
\label{sec.magvolt}

\begin{figure}[tbh]
\includegraphics[width=3.20in,clip]{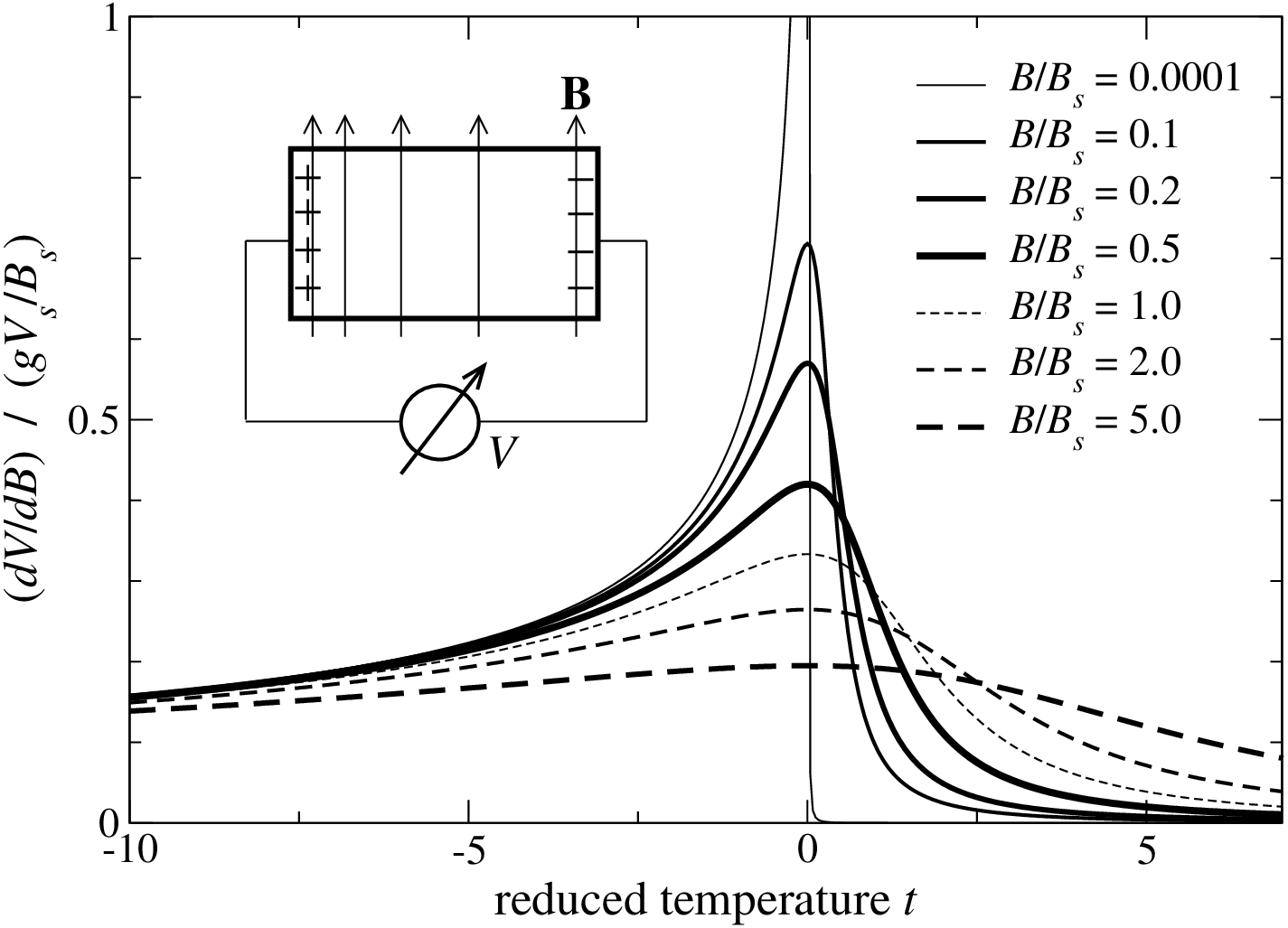}
\caption{\label{fig.diffVBB} Linear response of the voltage for a weakly
nonuniform magnetic field, $\partial V/\partial B\propto m_z\,\partial
m_z/\partial B$, as a function
of reduced temperature for various magnetic fields, assuming positive carrier
charge. Scaling factors and reduced
temperature are defined in Eqs.~(\ref{1.ms1})--(\ref{1.redD1}). Inset: Sketch of
DMS sample in a nonuniform field.}
\end{figure}

Let us consider the setup sketched in the inset in Fig.~\ref{fig.diffVBB}: A
metallic DMS sample is placed in an inhomogeneous magnetic field and the
voltage between opposite surfaces is measured. We assume the temperature to be
close to $T_c$ so that the magnetization is far from saturation. The
magnetization is larger in the region with strong magnetic field. This region
thus \emph{attracts} carriers due to the mechanism discussed above. An electric
current will only flow for a short time until the accumulated charge produces
an electric field that prevents further charge accumulation. For positive
(negative) carrier charge the electric potential in the strong-field region
will be higher (lower) than in the weak-field region. The voltage measured
between the two regions divided by the magnetic field difference will thus have
the same sign as the carrier charge.

Following Ref.~\onlinecite{Tim06} we start from a Landau theory for the coupled
magnetization and carrier density, defined by the Hamiltonian $H=H_m+H_{\delta
n}$ with
\ba
H_m & \!=\! & \int \! d^3r\, \Big\{ \frac{\alpha}{2}\,m^2
  + \frac{\beta}{4}\,m^4 + \frac{\gamma}{2}\, \partial_i\mathbf{m}
  \cdot\partial_i\mathbf{m} - Bm_z\! \Big\} ,\quad
\label{Hm1} \\
H_{\delta n} & \!=\! & \frac12 \int \! d^3r\,d^3r'\,
  \frac{e^2}{4\pi\epsilon_0\epsilon}\, \delta n(\mathbf{r})\,
  \delta n(\mathbf{r}')\, \frac{e^{-|\mathbf{r}-\mathbf{r}'|/r_0}}
  {|\mathbf{r}-\mathbf{r}'|} \nonumber \\
& & {}+ \int \! d^3r\, q\, \delta n(\mathbf{r})\, V .
\ea
The first part is the usual Landau functional for a Heisenberg ferromagnet in a
magnetic field $\mathbf{B}$ directed along the $z$ axis. Summation over
$i=x,y,z$ is implied. The second part
contains the screened Coulomb interaction and an
applied electric potential. The carriers are assumed to have charge
$q=\pm e$. We impose the constraint $\int d^3r\, \delta n(\mathbf{r}) = 0$.
The two parts are coupled through the
coefficient $\alpha$ in $H_m$, which depends on temperature and vanishes at the
Curie temperature. We write, to leading order,
$\alpha=\alpha'\,(T-T_c-\eta\,\delta n)$, where $T_c$ is now the Curie
temperature for $\delta n=0$.

The equilibrium states can be found from the Euler equations $\delta
H/\delta\mathbf{m}=0$ and $\delta H/\delta(\delta n)=0$.\cite{Tim06} We find
a paramagnetic phase for $T\ge T_c$, a uniform ferromagnetic phase for
$T^\ast\le T<T_c$, and a phase with periodic modulation of $\mathbf{m}$ and
$\delta n$ for $T<T^\ast$, where $T^\ast \equiv T_c -
{e^2\beta\gamma}/{\epsilon_0\epsilon{\alpha'}^3\eta^2}$.
By rescaling of length, magnetization, and energy it is possible to reduce the
number of parameters. For example, the magnetization can be written
in the scaling form
\be
\frac{\mathbf{m}(\mathbf{r})}{m_s} =
\mathbf{M}\!\left(g,t,\frac{B}{B_s};\frac{\mathbf{r}}{r_0}\right) ,
\ee
where
\be
m_s \equiv \frac{1}{r_0}\,\sqrt{\frac{\gamma}{\beta}} , \qquad
B_s \equiv \frac{1}{r_0^3}\,\sqrt{\frac{\gamma^3}{\beta}}
\label{1.ms1}
\ee
are the natural units of magnetization and magnetic field, respectively,
\be
t \equiv \frac{\alpha'(T-T_c)\, r_0^2}{\gamma}
\label{1.redt1}
\ee
is a dimensionless reduced temperature, and
\be
g \equiv \frac{\alpha'\eta\sqrt{\epsilon_0\epsilon}}{\sqrt{\beta}\,r_0 e}
\label{1.redD1}
\ee
is a dimensionless measure of the coupling $\eta$ between carrier concentration
and magnetization. $\mathbf{M}$ is a dimensionless vector scaling function.

\subsection{Uniform phase}

In the uniform ferromagnetic and paramagnetic phases, we can drop the gradient
term in $H_m$. We obtain the energy density $h=h_m+h_{\delta n}$ with
\ba
h_m & = & \frac{\alpha}{2}\, m^2 + \frac{\beta}{4}\,m^4 - Bm_z , \\
h_{\delta n} & = & \frac{e^2r_0^2}{2\epsilon_0\epsilon}\,
  \delta n^2 + q\, \delta n\, V 
  - q\, \delta n\, V_0 .
\ea
The last term implements the constraint of charge conservation with the Lagrange
multiplier $-qV_0$. From the Euler equation
$\partial h/\partial (\delta n)=0$ we obtain
\be
\delta n = -\frac{\epsilon_0\epsilon}{qr_0^2}\, (V-V_0)
  + \frac{\epsilon_0\epsilon}{e^2r_0^2}\, \frac{\alpha'\eta}{2}\,
  m^2 .
\ee
The constraint $\delta n=0$ leads to
\be
V-V_0 = \frac{\alpha'\eta}{2q}\, m^2 .
\label{1.V1}
\ee
Here, $V_0$ is a reference potential, which is irrelevant for the measured
voltages. From $\partial h/\partial\mathbf{m}=0$ together with $\delta n=0$ we
obtain $m_x=m_y=0$ and the standard result
$\alpha'(T-T_c) m_z + \beta m_z^3 - B = 0$.
The elementary solution for both $T<T_c$ and $T\ge T_c$ is
\be
\frac{m_z}{m_s}
  = \frac{-2\cdot 3^{1/3}\, t + 2^{1/3} (9\, b + \sqrt{81\, b^2
  + 12\, t^3})^{2/3}}{6^{2/3} (9\, b + \sqrt{81\, b^2
  + 12\, t^3})^{1/3}} ,
\label{1.mexact1}
\ee
where $b \equiv B/B_s$.
For weakly inhomogeneous magnetic fields we obtain the \emph{linear response}
of the voltage to the magnetic field gradient by expanding
Eq.~(\ref{1.V1}),
\be
\frac{\partial V}{\partial B} = \frac{\alpha'\eta}{q}\,m_z\,\frac{\partial m_z}
  {\partial B} .
\label{1.dVdB3}
\ee
An analytical expression can be obtained from Eq.~(\ref{1.mexact1}).
Using dimensionless quantities, we find
\be
\frac{\partial V}{\partial B}\, \left(\frac{V_s}{B_s}\right)^{\!-1}
  = {}\pm g\, \frac{m_z}{m_s}\,\frac{\partial}{\partial b}\,\frac{m_z}{m_s}
\label{1.dVdB4}
\ee
with $V_s \equiv \gamma/(r_0\sqrt{\beta\epsilon_0\epsilon})$. The sign is given
by the sign of the carrier charge. Results for different magnetic fields are
shown in Fig.~\ref{fig.diffVBB}. In the limit of vanishing magnetic field we
obtain, in the \emph{ferromagnetic} phase,
$\partial V/\partial B\, ({V_s}/{B_s})^{-1}
  = {}\pm ({g}/{2})\, {1}/{\sqrt{-t}}$.
The linear-response coefficient thus diverges as $T_c$ is approached
from below.

We next estimate the order of magnitude of $\partial V/\partial B$.
While the Landau theory does not apply down to $T=0$,
extrapolating to $T=0$ results in a magnetization of the correct order
of magnitude. For a rough estimate we thus write
\be
\left.\frac{\partial V}{\partial B}\right|_{B=0}
  \approx \frac{\eta}{2q}\,\frac{m_0(T\!=\!0)}{\sqrt{T_c(T_c-T)}} .
\ee
Here, $\eta\approx 5.4\times 10^{-25}\,\mathrm{Km}^3$ has been
estimated\cite{Tim06} from Ref.~\onlinecite{LLD05}, for the magnetization
we take the maximum value for $5\,\%$ Mn in (Ga,Mn)As, $m_0(T\!=\!0) \approx
5.13\times 10^4\,\mathrm{A}/\mathrm{m}$, and $T_c$ is set to $110\,\mathrm{K}$.
This gives a typical scale of ${\partial V}/{\partial B}|_{B=0}
\approx 7.9\times 10^{-4}\,\mathrm{V}/\mathrm{T}$ at low temperatures and the
coefficient increases like $(T_c-T)^{-1/2}$ as the temperature is increased.
This result suggests that the effect should be clearly measurable.

In the \emph{paramagnetic} phase, the magnetization $m$ is, to leading order,
linear in magnetic field. Thus the voltage, Eq.~(\ref{1.V1}), is quadratic in
$B$ and there is no linear response at zero field. Right at the critical point
the Landau-theory result for the magnetization is $m_z = (B/\beta)^{1/3}$,
leading to $V = ({\alpha'\eta}/{2q})\, (B/\beta)^{2/3}$.

\begin{figure}[tbh]
\includegraphics[width=3.20in,clip]{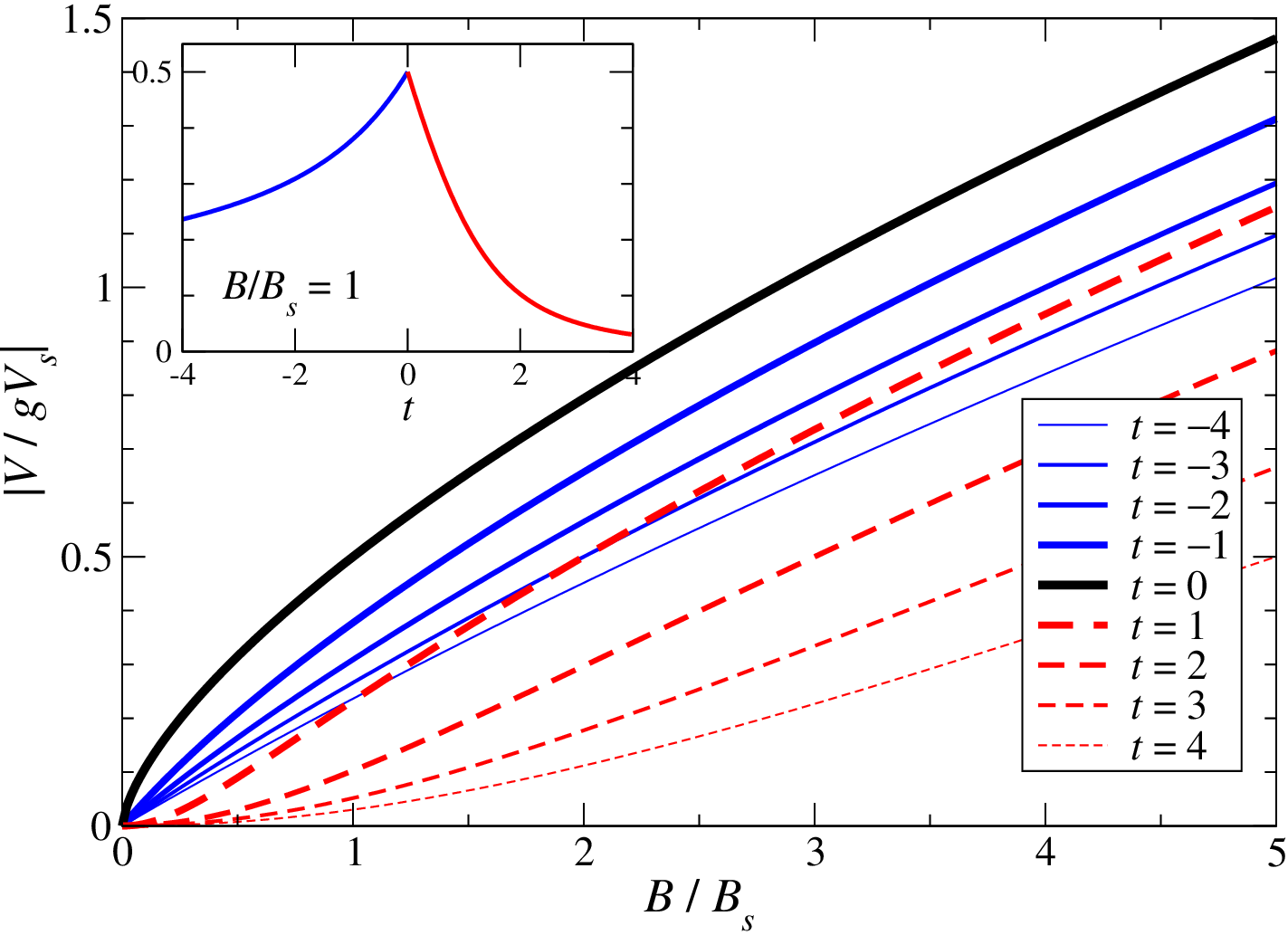}
\caption{\label{fig.VNL}(Color online) Magnitude of voltage $-V$ between
sample and ground as a function of magnetic field for various temperatures
for a DMS sample in the uniform ferromagnetic or paramagnetic phases.
Here, $t=\alpha'(T-T_c)r_0^2/\gamma$ is the dimensionless reduced temperature.
If the voltage is scaled with $1/g$, as it is plotted here, the curves do not
depend on $g$. Inset: Voltage as a function of $t$ for $B/B_s = 1$.}
\end{figure}

Next, we briefly discuss the voltage due to a \emph{large} change in magnetic
field. It may be possible to measure this voltage by first grounding a metallic
DMS sample in zero magnetic field, removing the ground connection, and then
applying a strong, uniform field. Since higher magnetization \emph{attracts}
carriers, a negative (positive) potential difference between sample and
ground is expected for positive (negative) carrier charge. Equation
(\ref{1.V1}) gives for vanishing magnetic field
$-V_0 = (\alpha'\eta)/(2q)\, m_0^2$,
where $m_0$ is the (uniform) magnetization in zero magnetic field,
which is $m_0 = \sqrt{-\alpha'(T-T_c)/\beta}$ in the ferromagnetic phase and
$m_0=0$ in the paramagnetic phase. Consequently,
\be
V = \frac{\alpha'\eta}{2q}\, (m^2 - m_0^2) .
\label{1.Vm1}
\ee
$V$ is the potential that would be necessary to maintain $\delta n=0$
if the sample were still connected to a charge reservoir.
If the sample is isolated, a voltage $-V$ with the reverse sign is measured.
In terms of dimensionless quantities we obtain
\be
-\frac{V}{V_s} = {}\mp\frac{g}{2}\,\left( \frac{m^2}{m_s^2}
  - \frac{m_0^2}{m_s^2} \right) ,
\label{1.Vm2}
\ee
where the upper (lower) sign applies to positive (negative) carrier charge. For
small magnetic fields we get the same linear-response result as above.
The full nonlinear response is obtained by inserting
Eq.~(\ref{1.mexact1}) into Eq.~(\ref{1.Vm1}). The resulting voltage is plotted
as a function of magnetic field for various temperatures in Fig.~\ref{fig.VNL}.
In the ferromagnetic phase, the voltage crosses over from a linear $B$
dependence to $B^{2/3}$ at $B/B_s \sim t^{3/2}$, see Eq.~(\ref{1.mexact1}).
In the paramagnetic
phase there is a crossover from $B^2$ to $B^{2/3}$ at the same scale. The inset
in Fig.~\ref{fig.VNL} shows the dependence of $|V|$ on temperature at a finite
magnetic field. The cusp at $T=T_c$ stems from the \emph{zero-field}
magnetization $m_0$ in Eq.~(\ref{1.Vm1}).

All results up to this point have been obtained from Landau mean-field theory.
The question arises whether the results
survive if fluctuations are
taken into account. After all, the average square magnetization $\overline{m^2}$
becomes an \emph{analytical} function of temperature through $T_c$.

We restrict ourselves to temperatures $T\approx T_c$ and to small magnetic
fields. Let us assume that charge fluctuations are fast compared to
fluctuations in the magnitude of $\mathbf{m}$. This seems reasonable since
charge density fluctuations have a typical timescale of the inverse plasma
frequency. In this case the carrier concentration instantaneously follows the
magnetization fluctuations. Equation (\ref{1.dVdB3}) then becomes
\be
\frac{\partial V}{\partial B} = \frac{\alpha'\eta}{2q}\,
  \frac{\partial \overline{m^2}}{\partial B} .
\label{1.dVdB6a}
\ee
Equation (\ref{Hm1}) shows that the averages of $m_z$ and
$m^2$ are obtained from the \emph{exact} free energy density $f$ by
$\overline{m_z} = -\partial f/\partial B$ and
$\overline{m^2} = 2\,\partial f/\partial \alpha$.
This implies the Maxwell relation
\be
\frac{\partial\overline{m^2}}{\partial B} = -2 \frac{\partial \overline{m_z}}
  {\partial\alpha} .
\ee
For vanishing magnetic field and $T<T_c$, one finds $\overline{m_z}\propto
(T_c-T)^{\underline{\beta}}$, where $\underline{\beta}$ is the usual critical
exponent (not to be confused with
the coefficient $\beta$ in the Landau functional) and $T_c$ is the \emph{true}
Curie temperature, taking fluctuations into account. Since $\alpha$ is a
linear function of $T$ we obtain
\be
\left.\frac{\partial V}{\partial B}\right|_{B=0}
  \propto (T_c-T)^{\underline{\beta}-1} .
\ee
This quantity still diverges as $T_c$ is approached from below. Note that for
the three-dimensional Heisenberg model, $\underline{\beta}\approx
0.37$.\cite{CHP02} Fluctuations thus make the divergence \emph{stronger}.
In the paramagnetic phase, $\overline{m_z}$ of course
vanishes for $B=0$ and we still do not find a linear response.

\begin{figure}[tbh]
\includegraphics[width=3.20in,clip]{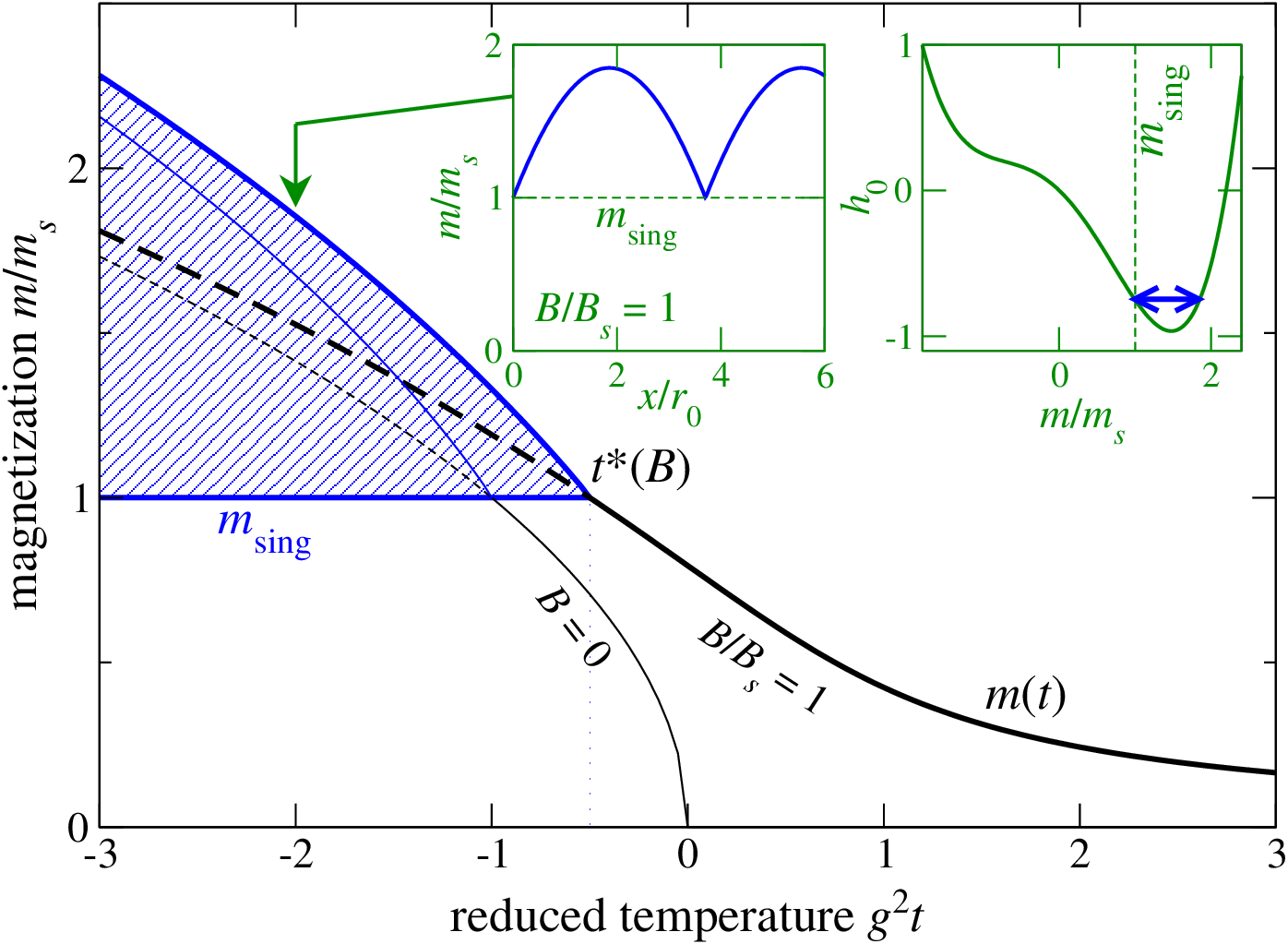}
\caption{\label{fig.hsol}(Color online) Diagram showing the equilibrium
phases as a function of reduced temperature
$g^2t$ in a nonzero magnetic field $B/B_s=1$.
For $t>t^\ast=\alpha'(T^\ast-T_c)r_0^2/\gamma$ the equilibrium state
is uniform with the magnetization given by the heavy solid line. For $t<t^\ast$
the magnetization shows periodic modulations spanning the cross-hatched
magnetization interval. The lower limit of the modulations is
$m=m_{\mathrm{sing}}$. The heavy dashed line shows the magnetization of the
uniform solution for $t<t^\ast$, which exists but has higher energy. The thin
lines denote the uniform magnetization and upper limit of modulations for
$B=0$, for comparison.
The left inset shows the magnetization as a function of position for
$g^2t=-2$ and $B/B_s=1$.
The right inset shows the effective potential $h_0$ as a function of
magnetization for the same parameter values. The double-headed arrow
denotes the magnetization modulation. The detailed numerical values depend
on the choice of $g$ ($g=1$ in this plot) and of $B/B_s$,
but the topology of the diagram does not.}
\end{figure}

\subsection{Stripe phase}

Below $T^\ast$ the Landau theory predicts an equilibrium state with periodic
modulations of magnetization and carrier concentration.\cite{Tim06} We
briefly review the pertinent properties of this state. The first two terms in
the Landau functional $H_m$, Eq.~(\ref{Hm1}), can be interpreted as a potential
for a fictitious particle with coordinates $\mathbf{m}$. The uniform
equilibrium states are determined by the minima of this potential.
In the stripe phase the coefficients are renormalized,\cite{Tim06} leading to
the energy density
$h_0 = ({\tilde\alpha}/{2})\,m^2 + ({\tilde\beta}/{4})\,m^4$
with
\ba
\tilde\alpha & \equiv & \alpha'(T-T_c)
  + \frac{{\alpha'}^2\eta^2\epsilon_0\epsilon}
  {2e^2r_0^2}\, \overline{m^2} , \\
\tilde\beta & \equiv & \beta
  - \frac{{\alpha'}^2\eta^2\epsilon_0\epsilon}{2e^2r_0^2} ,
\ea
where $m^2$ has to be
calculated selfconsistently. It turns out that a modulation of $m$ about the
minimum of $h_0$ is stable and energetically favorable for
$m>m_{\mathrm{sing}}$, where $m_{\mathrm{sing}} \equiv \sqrt{\gamma}
e/\alpha'\eta\sqrt{\epsilon_0\epsilon} = m_s/g$. Modulations occur between two
magnetizations with equal potential $h_0(m)$. For weak coupling, i.e., small
$\eta$ or $g$, $m_{\mathrm{sing}}$ is larger than the magnetization at the
minimum of $h_0(m)$ and stable modulations are not possible. For larger
coupling, $m_{\mathrm{sing}}$ is smaller than $m$ at the minimum, allowing
modulations to occur. The minimum-energy solution is obtained
for the \emph{maximum} possible modulation amplitude. In this case the lower
turning point of the modulation approaches $m_{\mathrm{sing}}$.\cite{Tim06}

\begin{figure}[tbh]
\includegraphics[width=3.20in,clip]{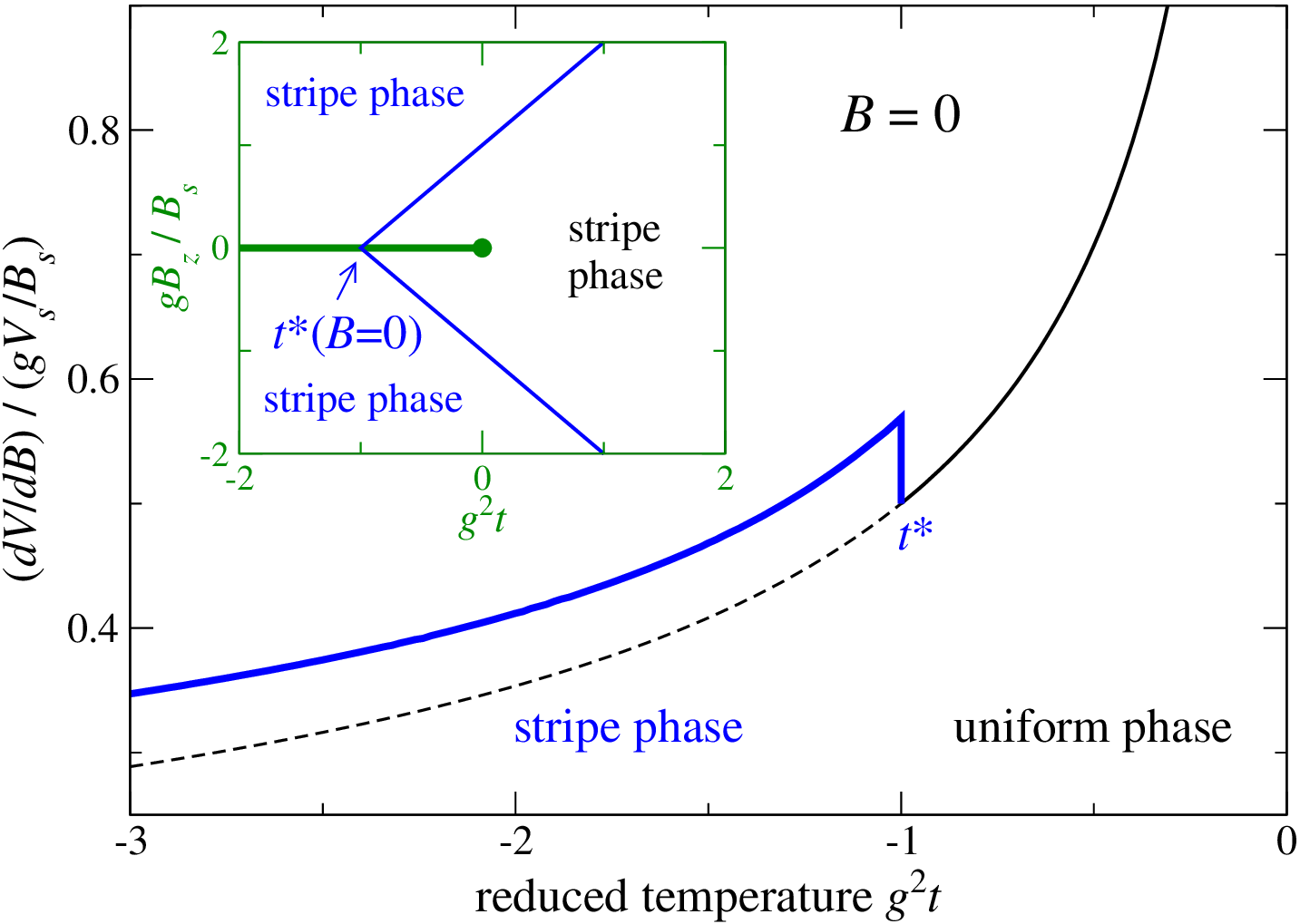}
\caption{\label{fig.dVdB}(Color online)
Linear response $\partial V/\partial B$ for small magnetic field,
as a function of reduced
temperature. The jump in $\partial V/\partial B$ at $t=t^\ast$ is
discussed in the text. For $t<t^\ast$, the result for the unstable uniform
state would continue along the dashed line.
The numerical calculation has been performed for the coupling $g=1$.
Inset: Phase diagram in the temperature--magnetic-field plane.
The heavy solid line denotes the first-order transition of the
ferromagnet, ending in a critical point at $T_c$ (i.e., $t=0$).
The thin solid lines emanating from $t=t^\ast(B\!=\!0)$ denote the
second-order transition between uniform and stripe phases.}
\end{figure}

In the present case we add the Zeeman term $-Bm_z$ to the energy density, which
tilts the effective potential, as shown in the right inset in
Fig.~\ref{fig.hsol}.
Numerical integration shows that at any temperature the modulation
with maximum amplitude still has the lowest energy. Figure \ref{fig.hsol} shows
a typical phase diagram in the temperature--magnetization plane. A typical
minimal-energy solution is shown in the left inset.

The field-dependent transition temperature $T^\ast(B)$ to the stripe phase is
determined by the condition $m=m_{\mathrm{sing}}$, where $m$ is the uniform
magnetization, Eq.~(\ref{1.mexact1}). The solution is
\be
T^\ast = T_c - \frac{\beta\gamma e^2}{{\alpha'}^3\eta^2\epsilon_0\epsilon}
  + \frac{\eta \sqrt{\epsilon_0\epsilon}}{\sqrt{\gamma}\,e}\, B
  = T^\ast(B\!=\!0)
  + \frac{\eta \sqrt{\epsilon_0\epsilon}}{\sqrt{\gamma}\,e}\, B .
\ee
Using Eqs.~(\ref{1.ms1})--(\ref{1.redD1}), this is equivalent to
\be
t^\ast \equiv \frac{\alpha'(T^\ast-T_c)r_0^2}{\gamma} = -\frac{1}{g^2}
  + g\,\frac{B}{B_s} .
\ee
Thus the transition temperature depends linearly on $B$ for arbitrary
$B$. Of course, for large magnetic fields Landau theory becomes inapplicable.
Note that while there is no sharp paramagnet-ferromagnet transition in nonzero
field, the transition to the stripe phase remains sharp. The reason is that
translational symmetry is preserved by the applied field and thus can be
spontaneously broken at this transition. The inset in Fig.~\ref{fig.dVdB} shows
the resulting phase diagram.


The analogue of Eq.~(\ref{1.V1}) in the stripe phase is
\be
V-V_0 = \frac{\alpha'\eta}{2q}\, \overline{m^2} .
\label{2.V2}
\ee
If fluctuations are neglected, the average is a spatial average over the
static modulation. The linear response $\partial V/\partial
B=(\alpha'\eta/2q)\,\partial \overline{m^2}/\partial B$ is
calculated numerically and the result for $B=0$ is shown in Fig.~\ref{fig.dVdB}.
The linear-response coefficient shows a
\emph{discontinuity} at $T^\ast$ within Landau theory. This phenomenon is
accompanied by a similar jump in the susceptibility at $T^\ast$ (not shown).
Physically, the system is more easily polarizable in the stripe phase, since
it has an additional parameter, the wavelength $\lambda$, that can be adapted
to the applied field. In agreement with this interpretation, the inset in
Fig.~\ref{fig.dVdB} shows that the stripe phase is \emph{stabilized} by a
magnetic field.


\section{Thermopower}
\label{sec.thermop}

Since the magnetization changes with temperature and is coupled to the carrier
concentration, there should also be a change in carrier
concentration or potential with temperature. In the case of an electrically
isolated sample one expects a nonzero thermopower
\be
Q \equiv - \left.\frac{\partial V}{\partial T}\right|_{I=0} .
\label{3.thermo1}
\ee
Consider the following setup: One end of a DMS sample is kept at a temperature
$T_1$, the other at temperature $T_2>T_1$. A discussion of the standard origin
of the thermopower can be found in textbooks.\cite{AM} The result is that the
$Q$ has the same sign as the carrier charge $q$. For this reason the
thermopower is often meassured to obtain the sign of this charge in
semiconductors, including DMS.\cite{HSY02,CCC02}


Here, we predict an additional contribution in DMS: In the same setup,
with $T_1 < T_2 < T_c$, the magnetization $m$ is larger in the cool region.
Carriers are attracted to high-magnetization regions, leading to a current
which results in the accumulation of positive (negative) charge in the cool
region for positively (negatively) charged carriers. This charge generates an
electric field that prevents further charge flow. The electric potential $\phi$
is higher (lower) in the cool region for positive (negative) carrier charge
$q$. With the explicit minus sign in Eq.~(\ref{3.thermo1}) this leads to a
thermopower of the \emph{same sign} as the normal contribution. Proceeding as
in Sec.~\ref{sec.magvolt} we obtain
\be
V = \phi(T_2) - \phi(T_1) = \frac{\alpha'\eta}{2q}\,
  \left[\overline{m^2}(T_2) - \overline{m^2}(T_1)\right] .
\label{3.DV2}
\ee
We restrict ourselves to vanishing magnetic field. In the uniform phase we can
write, neglecting fluctuations, $\overline{m^2}=m^2$.
In Landau theory, Eq.~(\ref{3.DV2}) gives
\be
Q \equiv Q_{\mathrm{uni}} = \frac{{\alpha'}^2\eta}{2q\beta} .
\label{3.thermo2}
\ee
Thus the thermopower is \emph{independent of temperature}
for $T^\ast<T<T_c$. On the other hand, for $T\ge T_c$ there is no magnetization
and this contribution to the thermopower vanishes. We thus find a jump at $T_c$.

\begin{figure}[tbh]
\includegraphics[width=3.20in,clip]{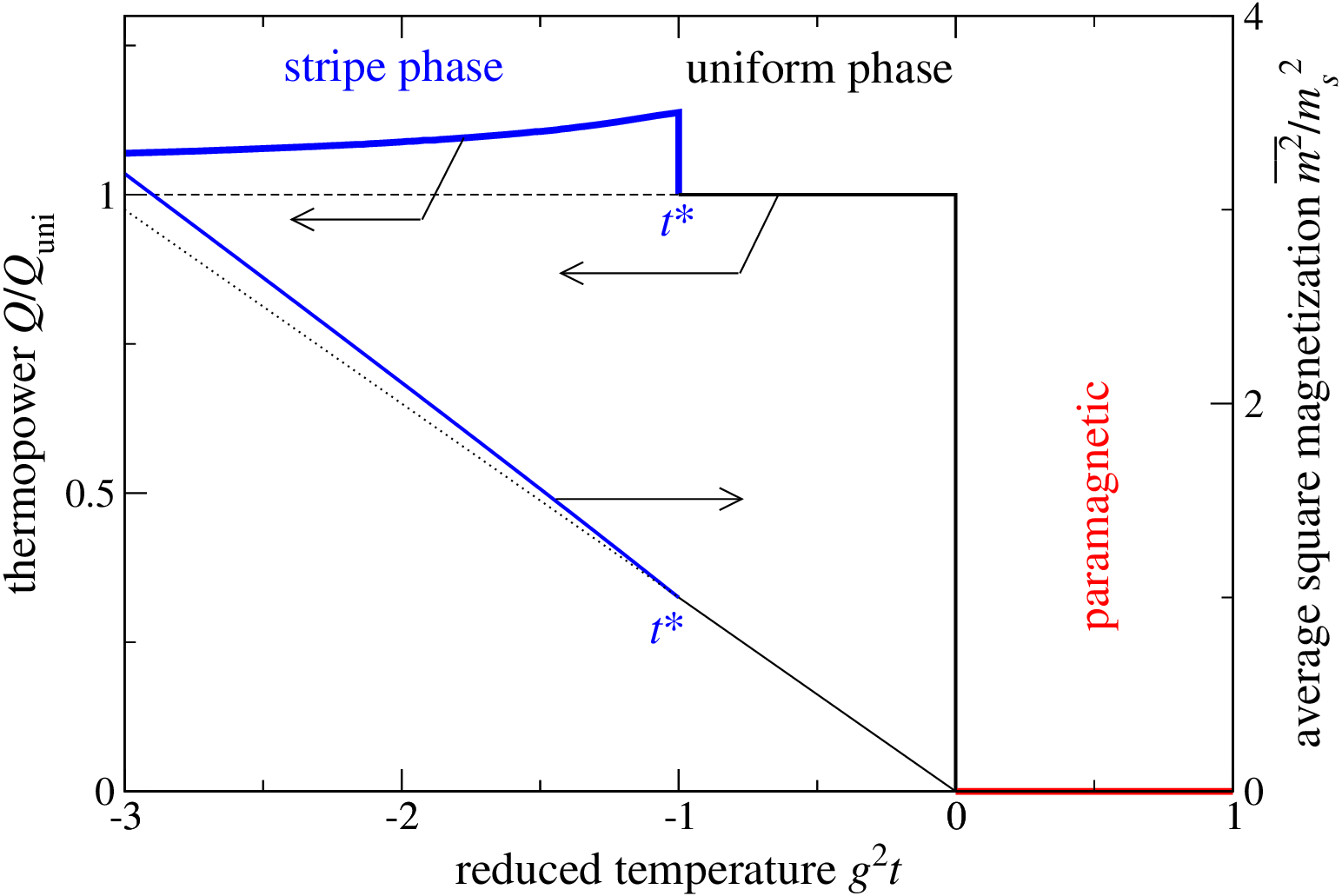}
\caption{\label{fig.thermoQ}(Color online) Left axis:
Thermopower $Q$ in units of the constant thermopower $Q_{\mathrm{uni}}$
in the uniform phase for vanishing magnetic field, as a function of
reduced temperature. Also shown is the average square magnetization
$\overline{m^2}$ (right axis). The numerical calculation for $t<t^\ast$ has
been performed for $g=1$.}
\end{figure}

In the stripe phase, for $T<T^\ast$, we find numerically that the thermopower
increases in magnitude with temperature. At $T^\ast$ it shows a downward jump,
since the average square magnetization $\overline{m^2}$ changes slope as a
function of temperature at the transition. Thermopower and $\overline{m^2}$ are
shown in Fig.~\ref{fig.thermoQ}.

To estimate the order of magnitude of the thermopower, we again replace
$\alpha'/\beta$ by $m^2(T\!=\!0)/T_c$, leading to
$Q=-\beta\eta m^4(T\!=\!0)/2qT_c^2$.
The only additional parameter needed beyond those given in
Sec.~\ref{sec.magvolt} is $\beta$. By comparing the mean-field result for the
gain in energy density due to magnetic ordering of the Heisenberg model on a
simple cubic lattice to the corresponding gain for the Landau theory, we obtain
\be
\beta \approx \frac{1}{(g\mu_B)^4}\, \frac{6k_BT_c}{S^3(S+1) n_{\mathrm{Mn}}^3}
  \approx 2.1\times 10^{-12}\,\mathrm{Jm}/\mathrm{A}^4  ,
\ee
where $g\approx 2$ is the g-factor for the impurity spins with $S=5/2$ and
$n_{\mathrm{Mn}}$ is the concentration of impurities, again assumed to be
$5\,\%$ of cation sites. This gives $Q \approx 2.0\times
10^{-3}\,\mathrm{V}/\mathrm{K}$.
In nonmagnetic semiconductors showing thermally activated conduction, the
magnitude of the thermopower is of the order of $10^{-5}$ to
$10^{-4}\,\mathrm{V}/\mathrm{K}$. In metals it is
a few times $10^{-6}\,\mathrm{V}/\mathrm{K}$.
Thus the thermopower generated by the carrier-magnetization
coupling is expected to be larger than the normal contribution.

The thermopower has been measured for various DMS,\cite{HSY02,CCC02} but
usually only to infer the sign, which is not changed by the physics discussed
here. A study of the thermopower in (Ga,Mn)As is under way.\cite{Shi05} It
would be worthwhile to look for an anomalously large thermopower in DMS.
However, the main signature of the magnetization-induced effect would be a
downward jump at $T_c$.

Finally, we briefly turn to the qualitative effect of fluctuations: The average
square magnetization becomes analytic in temperature through $T_c$ so that we
expect the jump in $Q$ to be replaced by a continuous crossover. Its width
should be given by the Ginzburg criterion for the fluctuation-dominated
temperature interval. In real DMS, disorder plays an important
role,\cite{Tim03} which may lead to additional broadening. However, for
high-quality metallic samples of (Ga,Mn)As, the magnetization curves close to
$T_c$ show a sharp decrease\cite{PKC01,EWC02,EWC02a,MSS03,KPW03} similar to the
Heisenberg model on a regular lattice, suggesting that disorder is not
dominant.

\section{Summary and Conclusions}
\label{sec.sum}

The strong coupling between carrier concentration and magnetization in
ferromagnetic DMS has been found to lead to a dependence of the electric
potential on the magnetization. If the magnetization is controlled by an
applied magnetic field, a magnetic-field-induced voltage is expected. Two
possible setups are discussed in this paper, which allow to study the linear
response for weakly nonuniform magnetic fields and the nonlinear voltage
induced by a strong field. The linear-response coefficient in the limit of
small field is estimated to be of the order of $10^{-3}\,\mathrm{V}/\mathrm{T}$
for (Ga,Mn)As at low temperatures. It diverges as $T_c$ is approached from
below. This singularity persists if fluctuations are taken into account.

On the other hand, if the magnetization is changed by varying the temperature,
the same physics leads to a variation of electric potential with temperature,
i.e., a thermopower. Within Landau theory, the thermopower is
temperature-independent in the uniform ferromagnetic phase and shows a
discontinuity at the Curie temperature $T_c$, which is smeared out by
fluctuations. For (Ga,Mn)As the estimate for this contribution to the
termopower is of the order of $10^{-3}\,\mathrm{V}/\mathrm{K}$, which is what
one would have guessed from the magnitude of the magnetic-field-induced
voltage. All these results are obtained for a uniform equilibrium
magnetization.

A nonuniform, stripe-like equilibrium magnetization is possible in DMS at lower
temperatures, $T<T^\ast$.\cite{Tim06} The phase diagram of uniform and stripe
phases is obtained within Landau theory. If the stripe phase were realized, the
magnetic-field-induced voltage would be qualitatively similar to the uniform
case, but the linear response $\partial V/\partial B$ would show a
discontinuity at $T^\ast$. The effect is larger in the stripe phase because the
magnetization can adapt more easily to the applied magnetic field by varying
the wavelength of the stripe pattern. The thermopower in the stripe phase is
temperature-dependent, unlike in the uniform phase, and shows a downward jump
at $T^\ast$.


\acknowledgments

The author thanks F. S. Nogueira for helpful discussions.

\end{document}